\documentclass[aps,twocolumn,showpacs,amsmath,amssymb,pre,superscriptaddress, floatfix]{revtex4}

\usepackage{graphicx}
\usepackage{dcolumn}
\usepackage{bm}
\usepackage{amssymb}
\usepackage{multirow}

\begin{document}

\title{Directed transport and localization in phase-modulated driven lattices}

\date{\today}


\author{Christoph Petri}
\email[]{Christoph.Petri@physnet.uni-hamburg.de}
\affiliation{Zentrum f\"ur Optische Quantentechnologien, Universit\"at Hamburg, Luruper Chaussee 149, 22761 Hamburg, Germany}%
\author{Florian Lenz}
\email[]{Florian.Lenz@physnet.uni-hamburg.de}
\affiliation{Zentrum f\"ur Optische Quantentechnologien, Universit\"at Hamburg, Luruper Chaussee 149, 22761 Hamburg, Germany}%
\author{Fotis K. Diakonos}
\email[]{fdiakono@phys.uoa.gr}
\affiliation{Department of Physics, University of Athens, GR-15771 Athens, Greece}
\author{Peter Schmelcher}
\email[]{Peter.Schmelcher@physnet.uni-hamburg.de}
\affiliation{Zentrum f\"ur Optische Quantentechnologien, Universit\"at Hamburg, Luruper Chaussee 149, 22761 Hamburg, Germany}%

\begin{abstract}
We explore the dynamics of non-interacting particles loaded into a phase-modulated one-dimensional lattice formed by laterally oscillating square barriers. Tuning the parameters of the driven unit cell of the lattice selected parts of the classical phase space can be manipulated in a controllable manner. We find superdiffusion in position space for all parameters regimes. A directed current of an ensemble of particles can be created through locally breaking the spatiotemporal symmetries of the time-driven potential. Magnitude and direction of the current are tunable. Several mechanisms for transient localization and trapping of particles in different wells of the driven unit cell are presented and analyzed.
\end{abstract}

\maketitle

\section{Introduction}

Time-driven systems represent a major focus in several versatile research fields, such as the physics of atoms, molecules or mesoscopic systems \cite{Gavrila:1992,Delone:1995,Krausz:2000}. Generally, they are evoked by the occurrence of time-periodic forces. As an example, the dipole interaction of atoms exposed to laser fields gives rise to several interesting phenomena like laser stabilization, high harmonic generation or above barrier ionization. Another example is the coherent control of quantum molecular dynamics by means of shaped femtosecond laser pulses.

Among the most prominent mesoscopic systems are the various driven lattice setups, i.e. particles in an one-dimensional static potential are acted upon additionally by external time-dependent forces of zero mean. Experimentally they have been realized in condensed matter \cite{Linke:1999,Majer:2003} and cold atomic systems (see \cite{Schiavoni:2003,Gommers:2005,Salger:2009} and Refs. therein). A remarkable observation is that these systems can show directed transport for an ensemble of particles, although there exists no net force. Therefore, they are called ratchets. Originally the generation of ratchet effects has been addressed by employing external noise \cite{Reimann:2002,Astumian:2002,Goychuck:2001}. Similarly, the role of dissipation has been studied thoroughly. In Ref. \cite{Haenggi:1996} it has been shown that directed transport occurs for underdamped particles in a sinusoidally rocked spatially asymmetric periodic potential. Moreover, as the amplitude of the external driving is varied, the current flow is reversed several times. The underlying mechanism responsible for the existence of directed currents and the reversal of the transport has been identified in Refs. \cite{Mateos:2000,Mateos:2003}. Due to dissipation transporting attractors in phase space emerge and by choosing the initial conditions appropriately it is possible to populate them selectively. The ratchet effect can also be generated in systems without dissipation and noise. In this case one speaks of ``(deterministic) Hamiltonian ratchets''. Recently, in Ref. \cite{Salger:2009} the directed transport of atoms in a BEC loaded into a flashing ratchet potential was demonstrated. The current flow appears only if certain temporal and spatial symmetries of the driven potential, which have been identified in Ref. \cite{Flach:2000}, are broken. The occurrence of directed transport in Hamiltonian systems has been reported for the case of fully chaotic dynamics \cite{Monteiro:2002,Hutchings:2004,Brumer:2006} and for systems with mixed phase space \cite{Gong:2004,Casati:2007,Denisov:2001,Denisov:2002}. For the latter case \cite{Schanz:2001,Schanz:2005,Dittrich:2000} a sum rule for the transport velocity of a classical ensemble of particles has been derived. Additionally, these authors have shown that in the semi-classical limit the quantum and the classical transport velocity coincide. As an extension in Ref. \cite{Denisov:2007} the impact of avoided crossing between different transporting Floquet states has been considered. Tuning the control parameters leads to an enhancement or suppression of the current flow. Moreover, in Ref. \cite{Denisov:2006} the influence of an additional dc bias on the directed transport of a Hamiltonian ratchet has been studied. The authors have found the persistence of transporting invariant submanifolds like regular islands. In their vicinity trajectories can get sticky, such that they perform ballistic-like motion. Remaining chaotic trajectories are accelerated by the bias field getting separated very fast from the ballistic type  dynamics.

For all previous setups the static potential is exposed to a so-called global driving law \cite{Gommers:2005,Salger:2009,Flach:2000,Denisov:2001,Denisov:2002,Schanz:2005}, i.e. the force acting upon the particles can be separated into two parts, which depend only on the spatial coordinate and the time, respectively. For the systems we explore in the present work this is not true anymore. Each potential barrier of the underlying lattice will be equipped with its own characteristic driving law, which gives rise to several new intriguing phenomena and the possibility to ``locally engineer'' the classical phase space. By adjusting carefully the parameters of the barriers and accordingly their driving laws specific parts of the phase space can be manipulated in a controllable manner, whereas the remaining portion stays mainly unaffected. Moreover, the symmetries derived in \cite{Flach:2000} can be broken by imposing spatially dependent phase shifts to the driving laws of the barriers, which will be called a ``phase-modulated lattice'' in the following. Thereby, a directed transport of an ensemble of particles is evoked and both the direction and the magnitude of the current flow can be tuned easily. Importantly, for specific ranges of the barriers' potential height the particles show different localization behavior depending on their location within the unit cell of the lattice.

The paper is organized as follows. In Sec. \ref{ch:setup} we give a detailed description of our model and define the setups which will be studied. In Sec. \ref{ch:pss} the phase space is analyzed by means of stroboscopic Poincar\'{e} surfaces of section (PSS). In this regard the desymmetrization of the phase space, meaning the loss of symmetry with respect to $p=0$, is explored. Furthermore, we study the occurrence of regular elliptic islands in the PSS, present an approximation for the velocity regime of the last stable torus and discuss the impact of cantori on the dynamics of trajectories. Section \ref{ch:trans_loc} is devoted to the transport and localization properties of the setups and how they originate from the phase space properties of the considered system. Finally, a conclusion and outlook is given in Sec. \ref{ch:sum}.


\section{Setup}\label{ch:setup}

\begin{figure}
\includegraphics[width=\columnwidth]{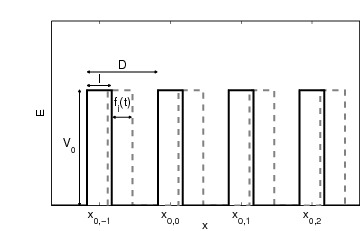}
\caption{\label{fig:fig1} Schematic illustration of the lattice of laterally oscillating barriers.}
\end{figure}

The classical dynamics of an ensemble of identical, non-interacting particles in an one-dimensional, infinitely extended lattice of laterally oscillating square potential barriers of equal height $V_0$ and width $l$ is described by the Hamiltonian
\begin{equation}\label{eq:ham}
H(x,p,t)=\frac{p^2}{2m}+V(x,t),
\end{equation}
where $m$ is the mass of the particles and
\begin{equation}
V(x,t)=\sum_{i=-\infty}^\infty V_0 \Theta \left(\frac{l}{2}-\left| x-x_{0,i}-f_i(t) \right| \right)
\end{equation}
is the potential. $x_{0,i}$ is the equilibrium position of the {\it i}-th barrier. $f_i(t)$ is a function with period $T$, which is explicitly allowed to depend on the site {\it i}. In the following, we will call $f_i(t)$ the driving law of the {\it i}-th barrier. The Hamiltonian \eqref{eq:ham} is called to be unbiased, if the force $F=-dV(x,t)/dx$ averaged over space and time vanishes. This condition is fulfilled, when each barrier of the lattice is driven with a periodic driving law. Accordingly, a natural choice for $f_i(t)$ is a harmonic function. Therefore we set
\begin{equation}
f_i(t)=C \cos(\omega \, t+\varphi_i),
\end{equation}
where $\varphi_i \in [0,2\pi[$ is a local phase shift, which depends on the {\it i}-th barrier. Here we have chosen the same amplitude $C$ and frequency $\omega$ for all barriers. Furthermore, we choose for the equilibrium position of the potentials
\begin{equation}
x_{0,i}=i \, D \quad \text{with} \enspace D>0,
\end{equation}
such that the static counterpart of Hamiltonian \eqref{eq:ham} is an equally spaced lattice of identical barriers, i.e. in their equilibrium position they are centered around $\{0,\pm D, \pm 2D,\ldots \}$ (see Fig. \ref{fig:fig1}). Fig. \ref{fig:fig1} shows a sketch of the lattice including the relevant parameters. Only for very special choices of the phases, e.g. $\{\varphi_i=0\}$ for all $f_i(t)$, neighboring barriers have equal spatial distance at a given snapshot.
To avoid overlap of the barriers for arbitrary choices of $\{\varphi_i\}$, the distance between the barriers must obey $D \geq 2C+l$.

The focus of this work are the transport properties of the system described by the Hamiltonian \eqref{eq:ham}. Let us therefore discuss under which conditions directed transport can occur. In Ref. \cite{Flach:2000}, the authors have proven that the directed current vanishes if the equation of motions are invariant under certain transformations, i.e. if certain symmetries hold. These transformations change the sign of the velocity and as a result it is possible to construct for every trajectory a mirror image with the opposite sign for the velocity. In this case, averaging over a representative set of trajectories yields zero mean velocity, i.e. a vanishing directed transport. Such transformations reverse either time or spatial coordinates together with a constant shift. Thus they take on the appearance
\begin{equation}\label{eq:trans}
\begin{split}
T_a& \quad : \quad x \rightarrow -x+c_x \enspace , \enspace  t \rightarrow t+c_t,\\
T_b& \quad : \quad x \rightarrow x+c_x \enspace , \enspace  t \rightarrow -t+c_t.
\end{split}
\end{equation}
The equations of motion belonging to the Hamiltonian \eqref{eq:ham} are simply given by
\begin{equation}
m \, \ddot{x}=-\frac{\partial{V(x,t)}}{\partial{x}},
\end{equation}
For the absence of directed transport for symmetry reasons it is therefore necessary that the potential $V(x,t)$ is invariant under one of the transformations \eqref{eq:trans}. We remark that a vanishing current can also occur if the above symmetries do not hold, which is however a non-generic case and is encountered only for specifically chosen parameter values. In the case of the uniformly oscillating lattice
\begin{equation}
\varphi_i=\text{const}. \enspace \forall i
\end{equation}
actually two such transformations can be identified.
\begin{equation}
\begin{split}
T& \quad : \quad x \rightarrow -x \enspace , \enspace  t \rightarrow t+\pi,\\
T'& \quad : \quad x \rightarrow x \enspace , \enspace  t \rightarrow -t.
\end{split}
\end{equation}
Consequently, for the uniformly oscillating lattice no directed transport will occur. The same is true, if the local phase shifts of the barriers alternate, i.e.
\begin{equation}
\varphi_i=\begin{cases} \varphi_1 & \text{if {\it i} is even} \\ \varphi_2 & \quad \text{else}. \end{cases}
\end{equation}
In this case the transformation is given by
\begin{equation}
T \quad : \quad x \rightarrow x+D \enspace , \enspace t \rightarrow -t.
\end{equation}
In order to reduce the number of varying parameters, we set the frequency and the amplitude of the driving laws to $\omega=1$ and $C=1$. Moreover, the barriers' width and the equilibrium distance are fixed to $l=0.4$ and $D=4.4$, respectively. Finally, without loss of generality the mass of the particles is chosen $m=1$. We therefore remain with two parameters: the barrier height $V_0$ and the local phase shifts $\{\varphi_i\}$. Since our major interest are the lattice's transport properties, we focus on setups, for which the symmetries \eqref{eq:trans} are broken. This is implemented by spatially modulating the phase shifts $\{\varphi_i\}$ of the driving laws. In the following three different setups are studied in detail:

(a) Firstly, the potential height is fixed to $V_0=0.16$ and the sites are equipped by a linearly increasing phase, i.e. a {\it constant phase gradient} is chosen,
\begin{equation}
f_i(t)=\cos \left(t+\frac{i}{n} 2\pi\right)
\end{equation}
where the periods of the site-dependent phases (for reasons of brevity we call them ``phase periods'') are $n=1,3,6,10$. For $n=3$, the sequence of phase shifts $\{\varphi_i\}$ is $\{\ldots,0,\frac{2\pi}{3},\frac{4\pi}{3},0,\frac{2\pi}{3},\ldots \}$, whereas for $n=1$ a lattice of uniformly oscillating barriers is recovered. One can verify straightforwardly that it is not possible to define proper $c_t$ and $c_x$, such that the potential is invariant under one of the transformations \eqref{eq:trans} for phase periods greater than or equal to three. Hence a nonzero mean velocity could be expected for $n=3,6,10$.

(b) For the second class of setups we keep $V_0=0.16$ and the above phase gradient with period three setup is perturbed ({\it perturbed phase gradient}), i.e. the sequence of phase shifts becomes $\{\ldots,0,\frac{2\pi}{3} \pm \alpha,\frac{4\pi}{3},0,\frac{2\pi}{3} \pm \alpha,\ldots\}$ such that for $\alpha=0$ the equidistant phase gradient is recovered.

(c) A specific phase period three $\{\ldots,0,\frac{\pi}{10},\frac{3\pi}{10},0,\frac{\pi}{10},\frac{3\pi}{10},\ldots \}$ is chosen and the global potential height $V_0$ is varied ({\it broken phase gradient}).

\section{Analysis of phase space}\label{ch:pss}

This section is devoted to the analysis of the phase space of the Hamiltonian \eqref{eq:ham} for the above-provided setups. General characteristics of the PSS, like the appearance of regular islands and the chaotic sea, are discussed. Especially, main emphasis is placed on the comparison of the phase space properties of the uniformly oscillating lattice $n=1$ (global driving, $n=1$) and the setups with spatial phase modulation (local driving, $n\geq3$). Furthermore, the stability of the phase space structures against perturbation of the phase pattern is studied.

\begin{figure}
\includegraphics[width=\columnwidth]{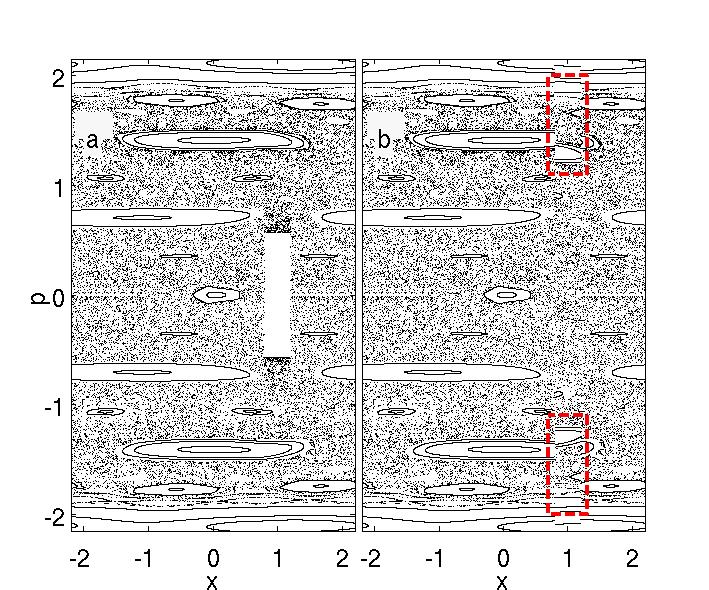}
\caption{\label{fig:fig2} (Color online) Stroboscopic PSS of the lattice of uniformly oscillating square potentials. For a better illustration we have added in (a) the potential energy to the kinetic energy of the particles, in order to avoid the discontinuities in the ``regular'' PSS (see (b)).}
\end{figure}

\subsection{Poincar\'{e} surfaces of section: Uniformly oscillating lattices}\label{ch:pss1}

In this section the phase space in the case of global harmonic driving is analyzed. Firstly, the method, how the Poincar\'{e} surfaces of section for this and all the upcoming setups are obtained, is described briefly. Afterwards, the phase space is discussed and a kinematic approach to determine the border of the chaotic sea is presented. Finally, the flux through cantori and the transit time of trajectories is determined.

Since the particles move at constant velocity between collisions with the barriers' edges, the Hamiltonian flow can be described by an implicit two-dimensional map (for details see \cite{Koch:2008}). To visualize the dynamics, we make Poincar\'{e} surface of sections (PSS) of the phase space. Due to the time-periodicity of the Hamiltonian $H(x,p,t+T)=H(x,p,t)$ ($T=2\pi$), an area-preserving surface of section is obtained by taking stroboscopic ``snapshots'' of the $(x,p)$-plane at times $t=n \, T$ with $n \in \mathbb{N}$. Additionally, the Hamiltonian possesses the translation invariance $H(x+L,p,t)=H(x,p,t)$, where $L$ is the length of the system's unit cell. Generally, $L$ is not the distance between the barriers at their equilibrium positions, but for our chosen setups $L$ is a multiple of $D$, i.e. $L=n \, D$, where $n$ is the phase period. Thereby, the space coordinate can be restricted to the length $L$ of the Hamiltonian's unit cell. The desired Poincar\'{e} surface of section is then provided by the set of points
\begin{equation}
\mathcal{M}=\{(x(t+kT)\bmod \, L,p(t+kT))|k \in \mathbb{N}\}.
\end{equation}
Obviously, since the time between successive ``snapshots'' of the $(x,p)$-plane is equal, the PSS is done for a specific phase of the driving. It is therefore not surprising that the appearance of the surface of section is not universal but depends on exactly this phase. An example of how this manifests itself in the PSS is given after the next paragraph.

In Fig. \ref{fig:fig2} (b) the PSS of the lattice with a phase period one, i.e. a lattice of uniformly oscillating barriers ($n=1$,$L$=$D$) is shown. This Poincar\'{e} surface of section possesses discontinuities (dashed box in Fig. \ref{fig:fig2} (b)), which are due to the fact that the particles are either inside or outside the barrier when the snapshot of the $(x,p)$-plane is taken. The particles outside have $E_{pot}=0$ and accordingly $E_{pot}=V_0$ inside. Since the potential is not smooth, this provides discontinuities in the PSS. For illustrative reasons it is therefore useful to add for those particles, which are at the moment of the snapshot inside the barrier, the potential energy $E_{pot}=V_0$ to the kinetic energy. This has been done in Fig. \ref{fig:fig2} (a). The invariant curves are then continuous, yet the barrier can be seen as a blank squared region in the PSS. Nevertheless, to avoid the discontinuities all the subsequent PSS are presented in this way.

Another eye-catching feature of the PSS is reflection symmetry with respect to $p=0$. However, this is not universally valid, but it occurs only for specific phases of the snapshot of the $(x,p)$-plane. As Fig. \ref{fig:fig2} (a) shows, the PSS is done at the moment the barrier arrives at one of its turning points, i.e. the {\it i}-th barrier is centered around $x_{0,i}=iD+C$ at this time instant. For the first unit cell $i=0$, this yields $x_{0,0}=1$ (blank squared region in Fig. \ref{fig:fig2} (a)). If the PSS is made for instance when the barrier has its maximum velocity, then this symmetry is absent. Yet, the dynamics of the lattice of uniformly oscillating barrier can be completely classified by means of the stroboscopic Poincar\'{e} surface of section \cite{Schanz:2005}.

\begin{figure}
\includegraphics[width=\columnwidth]{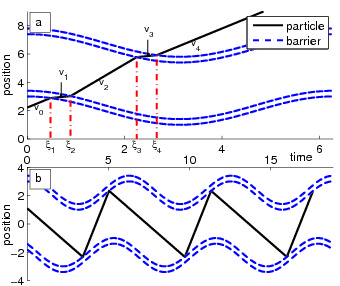}
\caption{\label{fig:traj} (Color online) (a) Central periodic orbit ($x\approx 2.1, \, p\approx 1.4$) for the resonance with winding number $w=2$ in the lattice of uniformly oscillating barriers. (b) Periodic two bounce orbit between the first and second barrier in the case of the specific phase period three $\{\ldots,0,\frac{\pi}{10},\frac{3\pi}{10},0,\frac{\pi}{10},\frac{3\pi}{10},\ldots \}$.}
\end{figure}

\begin{figure}
\includegraphics[width=\columnwidth]{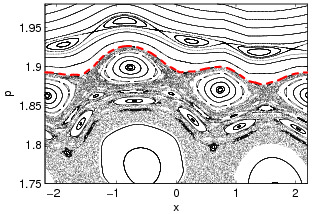}
\caption{\label{fig:fig2b} (Color online) Magnification of the Poincar\'{e} surface of section in the region of the upper boundary of the chaotic sea. The dashed curve is the last stable KAM torus, i.e. this invariant curve delimits the chaotic sea.}
\end{figure}

For large momenta $|p|\gg 1$ the potential is negligible due to its finite height. In this limit, the integrable dynamics of a free particle is recovered and the phase space is foliated by invariant curves, which are topologically equivalent to a torus. In the PSS these curves appear as straight lines, which stretch out over the whole unit cell. This simple part of the PSS is not displayed in Fig. \ref{fig:fig2} (a) and occurs for $|p|\geq 3$. With decreasing momentum, the integrability is lost, but large regular domains remain. Many orbits still lie on deformed tori, which occur as curved lines in the PSS ($|p| \approx 2$ in Fig. \ref{fig:fig2} (a)). Particles on these invariant curves travel through the lattice in the direction of their initial momentum. Indeed the KAM theorem predicts that tori with sufficiently irrational winding number survive under a small perturbation. The winding number is defined as the limit
\begin{equation}
w=\lim_{t \rightarrow \infty} \frac{x(t)-x(0)}{t},
\end{equation}
if it exists. $x$ and $t$ are measured in multiples of the spatial period of the unit cell and time period, respectively. Thus $w$ is proportional to the average velocity of the particle in the lattice $w \sim \overline{v}$. Tori with rational winding numbers $w=r/s$ are excluded in the KAM theorem. According to the Poincar\'{e}-Birkhoff theorem they dissolve into an even number of alternately elliptic and hyperbolic fixed points of period $s$. The trajectories in the extended system, which correspond to these periodic orbits in the PSS, travel $r$ spatial unit cells in $s$ time periods in the direction of their initial velocity. Hence in the PSS they occur as $s$ distinct points. Around each elliptic periodic orbit there is again a set of invariant curves, which can be seen in our PSS as elliptic islands. Completely analogous to the trajectory belonging to the periodic orbit in the center of this structure, the motion of particles, which are inside these islands proceeds through the lattice only in one direction. Accordingly, the trajectories intersect sequentially the PSS at different islands of the corresponding chain of islands. In the vicinity of the hyperbolic fixed points there is an infinite number of homo- and heteroclinic intersections of the stable and unstable manifolds, which yields horseshoe type dynamics and hence the presence of chaos. For large kinetic energies these chaotic layers are too small to be visible in the PSS. The magnification of the PSS (see Fig. \ref{fig:fig2b}) for $1.75<p<1.95$ shows such a separatrix region around $p \approx 1.93$. However, with decreasing momentum the strength of the perturbation and thereby the size of these layers increases until finally a large chaotic sea in the PSS for small momenta develops. Trajectories, belonging to this chaotic sea in the PSS, wander diffusively through the lattice.

At $p=0$, there is an elliptic island, which has been discussed in detail in \cite{Koch:2008}. It corresponds to trapped motion in the scattering region of a single barrier. Moreover, this island exists only for a certain regime of the parameters $(V_0,l)$. Due to the point-like interaction between the barrier and the particles it is possible to determine analytically \cite{Koch:2008} the position of these resonances in the PSS by exploiting the symmetry properties of the corresponding central periodic orbit. Other dominant elliptic islands in the PSS are the ones with winding number $w=r$, i.e. the central periodic orbits are trajectories, which travel $r$ lattice sites during one oscillation period of the lattice. Fig. \ref{fig:traj} (a) shows the trajectory, which corresponds to the central periodic orbit of the $r=2$ resonance at $x \approx 2.1, \, p\approx 1.4$ in Fig. \ref{fig:fig2} (a), together with the two barriers in position space. Obviously the phases belonging to the four collisions are symmetric with respect to $\frac{\pi}{2}$, ($\xi_4=\pi-\xi_1$, $\xi_3=\pi-\xi_2$), such that the velocity $v_4$ after transmission through the two barriers equals the initial velocity $v_0$. In the appendix \ref{ap:po_few} a method to determine the positions of periodic orbits with few collisions in the PSS is described using the example of this $w=2$-orbit.

Let us now discuss the border regions of the PSS ($1.75<p<2.0$ in Fig. \ref{fig:fig2} (a)), where the large chaotic sea is bounded by tori, which have not been destroyed by the perturbation, i.e. the driving. In this regime of intermediate kinetic energy more chains of elliptic islands are visible in the PSS. For even larger momenta the chaotic sea is bounded by an invariant curve, which is called the ``first invariant spanning curve'' (FISC). Fig. \ref{fig:fig2b} shows a magnification of this region of the PSS for positive momenta with the FISC shown as a dashed line. For negative momenta there exists of course a FISC, too (Fig. \ref{fig:fig2} (a)). Although the position of the FISC in the PSS is very hard to be determined analytically, one can derive at least an approximation to its position in velocity space. We recall that particles on the FISC are fast enough to be transmitted, even if a collision takes place when the barrier moves with its extremal velocity $u_{max}$ in the same direction, i.e. the minimal relative kinetic energy must exceed the potential height. Thus we get
\begin{equation}\label{eq:fis}
\frac{m\,(v-u_{ex})^2}{2}\geq V_0,
\end{equation}
where $v$ is the velocity of the particle. In case of a harmonic driving law, the extremal velocities of the barrier are given by $u_{ex}=\pm \omega C$. Plugging $u_{ex}$ in equation \eqref{eq:fis} yields an approximation to the velocity regime where the FISC is located in the PSS. Depending on the sign of $v$ we find
\begin{equation}\label{eq:fisc}
v_{\pm}= \pm \omega C \pm \sqrt{\frac{2}{m} V_0},
\end{equation}
where the index $\pm$ denotes the position of the FISC in the PSS for positive and negative velocities, respectively. In case of the harmonic driving law we have $|v_+|=|v_-|$ independent of the phase period. For $m=1$, $\omega=1$, $C=1$ and $V_0=0.16$, we find from equation \eqref{eq:fisc} $|v_{\pm}|\approx 1.57$. Fig. \ref{fig:fig2b} shows that the FISC of the uniformly oscillating lattice is located indeed at $p \approx 1.9 > m \cdot v_{+}$. This discrepancy is significant (around 15 \%), which is due to the fact that this approximation accounts only for a simple kinematic consideration of the dynamics between the particle and a single barrier and neglects all the dynamical processes happening in the extended system. Nevertheless, this naive approach provides some insights, as we shall argue briefly in the following. For driving laws with more than one frequency, which have been used for instance in Ref. \cite{Schanz:2005}, the extremal velocity $u_{ex}$ of the barrier depends on its sign. For example, in the case of the biharmonic driving law $f(t)=C(\sin(\omega t)+\cos(2 \omega t)$ one finds $u_{ex}=2\omega C$ and $u_{ex}=-3\omega C$, respectively. In this case equation \eqref{eq:fisc} yields $v_+=2 \omega C +\sqrt{2V_0/m}$ and $v_-=-3 \omega C - \sqrt{2V_0/m}$. Consequently, we expect that the FISC for positive momenta is located closer to $p=0$ than for negative momenta independently of the phase of the PSS. Indeed in Ref. \cite{Schanz:2005} a similar behavior has been observed.

\begin{figure}
\includegraphics[width=\columnwidth]{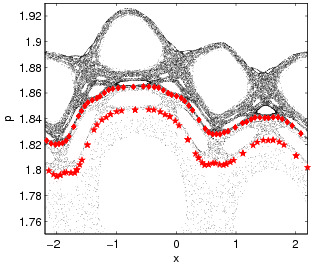}
\caption{\label{fig:cant} (Color online) Poincar\'{e} surface of section of a single chaotic trajectory, which has been started in the upper region and stopped once it has crossed the $p=0$-axis. The diamonds and stars are periodic orbits corresponding to truncations of the continued fraction expansion of $w_1=\frac{2\gamma+1}{\gamma}$ and $w_2=\frac{29\gamma-1}{11\gamma}$, respectively.}
\end{figure}

\subsection{Flux through cantori and transit times}

Let us now turn to the discussion of the transit times of orbits in the PSS. First of all we give an example of trajectories in the chaotic sea, which are confined to a subpart of phase space for a very long time. Then a procedure how to approximate the average escape time from this region is described briefly. Finally, this method is applied exemplarily in the case of the uniformly oscillating lattice.

In Fig. \ref{fig:cant} a magnification of the PSS of a single trajectory is shown. It has been launched in the chaotic layer close to the FISC and once it has crossed the {\it x}-axis, the simulation has been stopped. Obviously, there are sudden changes in the density of points in the PSS. This is a hallmark of the so-called cantori, which are remnants of dissolved tori with irrational winding number. These objects can be regarded as tori with gaps, so that the Hamiltonian flow is able to pass through \cite{Meiss:1992}. The diamonds and stars in Fig. \ref{fig:cant} are periodic orbits, which belong to truncations of the continued fraction expansions of cantori with the noble winding numbers $w_1=\frac{2\gamma+1}{\gamma}=[2,1^\infty]$ and $w_2=\frac{29\gamma-1}{11\gamma}=[2,1,1,2,1^\infty]$, where $\gamma=\frac{1+\sqrt{5}}{2}$ is the golden mean. The associated rational winding numbers for the two periodic orbits are
\begin{equation}
\begin{split}
w_\bullet&=[2,1^9]=\frac{144}{55}\\
w_\star&=[2,1,1,2,1^5]=\frac{129}{50}.
\end{split}
\end{equation}
As long as the trajectories are confined in the PSS to the region above the cantorus, their velocity never changes sign. Moreover, the magnitude of the velocity varies only over a narrow interval. Thus, they perform ballistic-like motion during this time span. This has severe impact on the dynamics of an ensemble of particles, which will be discussed in the next section. In order to obtain the average length of the ballistic flights, we have to determine the flux $\Phi_w$ through a cantorus with irrational winding number $w$. Mather \cite{Mather:1986} has shown that the sequence of differences in action of periodic orbits belonging to truncations of the continued fraction expansion of $w$ converges to $\Phi_w$
\begin{equation}
\Phi_w=\lim\limits_{\frac{r}{s} \rightarrow w}{\left(W_{r/s}-W^\ast_{r/s}\right)},
\end{equation}
where $W_{r/s}, \,W^\ast_{r/s}$ is the action of the ``minimizing'' and the ``minimax'' orbit respectively \cite{Meiss:1992}, i.e. periodic orbits with winding number $w=r/s$, belonging to the minimum (saddle point) of the action. The minimizing orbit is generically unstable and hyperbolic, whereas the minimax orbit is either an elliptic or hyperbolic-with-reflection periodic orbit. In the following $\Phi_{r/s}=W_{r/s}-W^\ast_{r/s}$ is called as the flux through a chain of periodic orbits with winding number $w=r/s$.
\begin{figure}
\includegraphics[width=\columnwidth]{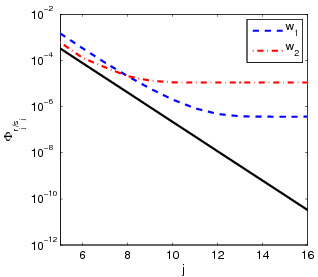}
\caption{\label{fig:flux} (Color online) Flux through the chains of periodic orbits, belonging to convergents of $w_1$ and $w_2$, as a function of the level $j$. The black curve is $\Phi_{r_j/s_j} \sim C \xi^{-j}$ with $\xi \approx 4.339$.}
\end{figure}
Due to the point-like interaction, the action $W_{r/s}$ of a periodic orbit can be calculated very easily. Between successive collisions with the barriers' edges, the particles move ballistically in a constant potential. Hence the Lagrangian is simply
\begin{equation}\label{eq:lagrangian}
L(x,\dot{x},t)=\begin{cases} \quad \frac{mv^2}{2} & \text{particle between barriers}\\ \frac{mv^2}{2}-V_0 & \text{particle in barrier} \end{cases}
\end{equation}
Consequently, the action of a periodic orbit in the PSS of period $s$ is given by
\begin{equation}
W_{r/s}=\sum_{i=0}^{s-1} \int_{t_i}^{t_{i+1}} L(x,\dot{x},t)dt
\end{equation}
$t_i=i 2\pi$ is the moment, when the {\it i}-th PSS is taken. Since the interaction is point-like this reduces to
\begin{align}\label{eq:action}
W_{r/s}=\sum_{i=0}^{s-1} \sum_{j=0}^{j_{\text{max}}(i)} W^j_{r/s} 
\end{align}
with
\begin{equation}
W^j_{r/s}=\begin{cases} \qquad \frac{mv^2}{2}\Delta t_{i,j} & \text{particle between bariers}\\ \left( \frac{mv^2}{2}-V_0 \right) \Delta t_{i,j} & \text{particle in barrier,} \end{cases}
\end{equation}
where $\Delta t_{i,j}=t_{i,j+1}-t_{i,j}$ are the discrete time intervals between the {\it i}-th and the {\it i}+1-st intersection of the trajectory with the surfaces of section, i.e. $t_{i,0}=i 2\pi$, $t_{i,j_{\text{max}}(i)}= (i+1) 2\pi$ and $t_{i,j(i)}$ are the points in time of the collisions occurring between the particle and one of the barriers' edges during $t_i$ to $t_{i+1}$. For the search of the periodic orbits a variational scheme has been adopted, which allows the detection of orbits with periods up to many thousands. In Appendix \ref{ap:po_many} a description of this method is presented. Fig. \ref{fig:flux} shows the flux $\Phi_{r_j/s_j}$ through convergents of $w_1$ and $w_2$ as a function of the level $j$ of the truncated continued fraction expansions. In the beginning the flux scales according to the power-law $\Phi_{r_j/s_j} \sim C \xi^{-j}$ with $\xi \approx 4.339$ \cite{MacKay:1983} (black curve in Fig. \ref{fig:flux}). Still the sequence converges rapidly with increasing {\it j} as Fig. \ref{fig:flux} shows. Asymptotically, we find for the fluxes through the cantori $\Phi_{w_1}=3.62 \cdot 10^{-7}$ and $\Phi_{w_2}=1.09 \cdot 10^{-5}$. An approximation to the transit time for particles to get from the region above the cantorus with winding number $w_i$ to the phase space below it is given by
\begin{equation}
t_{w_i}=\frac{A_{w_i}}{\Phi_{w_i}},
\end{equation}
where {\it A} is the area in the PSS above $w_i$.  By dividing the PSS into small squares, we have estimated these areas to be $A_{w_1}=0.12$ and $A_{w_2}=0.18$, which yields for the total transit time $t_{\text{transit}}=t_{w_1}+t_{w_2}=3.48 \cdot 10^5$ to get in the PSS from the region above the cantorus $w_1$ to the region below the cantorus $w_2$. Consequently, on average the particles get confined 348.000 periods of the driving in this region of the phase space and thus perform during this time ballistic flights. We note that the above transit time is meant to provide a rough estimate for the transport through the cantori. The comparison with numerical data obtained by simulating an ensemble placed in a chaotic region close to the FISC shows that the transit time is in general even longer. Beside the effect of being trapped to the region above the cantorus, the trajectories accessorily can get sticky to the hierarchy of elliptic islands surrounded by subislands being above the cantorus. The latter process extends the ballistic flights significantly.

\begin{figure*}
\includegraphics[width=2.0\columnwidth]{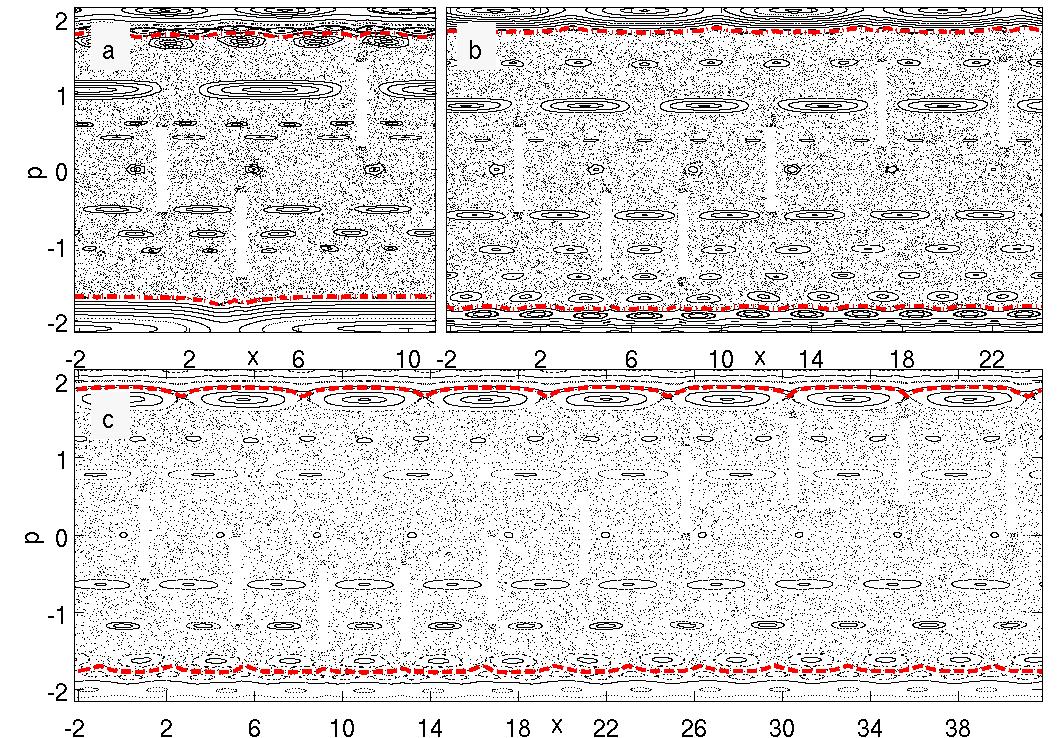}
\caption{\label{fig:fig3} (Color online) Stroboscopic PSS of the lattice with a linear phase gradient of period 3 (a), 5 (b) and (10). The length of the unit cell increases with the period of the gradient. The dashed curves are the FISCs.}
\end{figure*}

\subsection{Phase-modulated lattices}\label{ch:pss1}

So far we have considered only the lattice of uniformly oscillating barriers (global driving), i.e. $\{\varphi_i=\text{const}. \, \forall i\}$. Let us now turn to the setups, which have been defined in section \ref{ch:setup}. We will discuss especially the differences concerning their phase space structure compared to the setup with global driving. Firstly, the setup with constant phase gradient is studied. Then the stability of the phase space structure with respect to a perturbation of the gradient is considered. Finally, we study the impact of the potential height $V_0$ on the dynamics.

\paragraph{Constant phase gradient.} In Fig. \ref{fig:fig3} (a)-(c) the PSS of the lattice with a linear, equidistant phase gradient of period 3, 5 and 10 is shown. Contrary to the case $n=1$, the PSS is obviously not symmetric with respect to $p=0$ and moreover there is no alternative phase of a chosen PSS for which this symmetry is restored. As a result not only the PSS but the complete phase space of these phase periods is asymmetric with respect to $p=0$. In the next section we will see that this desymmetrization is actually the origin of the occurrence of directed currents in the system. Nevertheless, the PSS shows in general the same structure, which we have already discussed for the lattice of uniformly oscillating barriers. For small momenta there is a large chaotic sea with embedded resonances. Since the potential height $V_0$ and the potential width $l$ are equal for all phase periods, each barrier possesses analogous to $n=1$ at $p=0$ a small elliptic island of bounded motion in its scattering region \cite{Koch:2008} (e.g. $x_1 \approx 2, \, x_2 \approx 6, \, x_3 \approx 11$ for $n=3$ in Fig. \ref{fig:fig3} (a)). With increasing kinetic energy more chains of resonances appear until the chaotic sea is bounded by the FISCs. Although the form of the FISC depends on the phase at which the PSS is taken, Fig. \ref{fig:fig3} shows that for all periods of the gradient it is located in the velocity range defined by equation \eqref{eq:fisc}. However, a significant symmetry the phase space possesses for $n=1$ has disappeared, namely the areas of the elliptic islands in the PSS with winding numbers $w=r/s$ and $w=-r/s$ are usually not equal anymore. For example in the case $n=3$ the dominant resonance is the one with winding number $w=1/2$ ($p \approx 1, \, x \approx 0.5$ in Fig. \ref{fig:fig3} (a)). On the other hand there are three resonances with $w=-1/2$ ($p \approx -1,\, x_1 \approx 0.5, \, x_2 \approx 3, \, x_3 \approx 5$ in Fig. \ref{fig:fig3} (a)). Apparently, their total area in the PSS is less than the area of the single $w=1/2$ resonance. Since these areas are preserved under the Hamiltonian flow, this is universally valid for all phases of the PSS.

\begin{figure}
\includegraphics[width=\columnwidth]{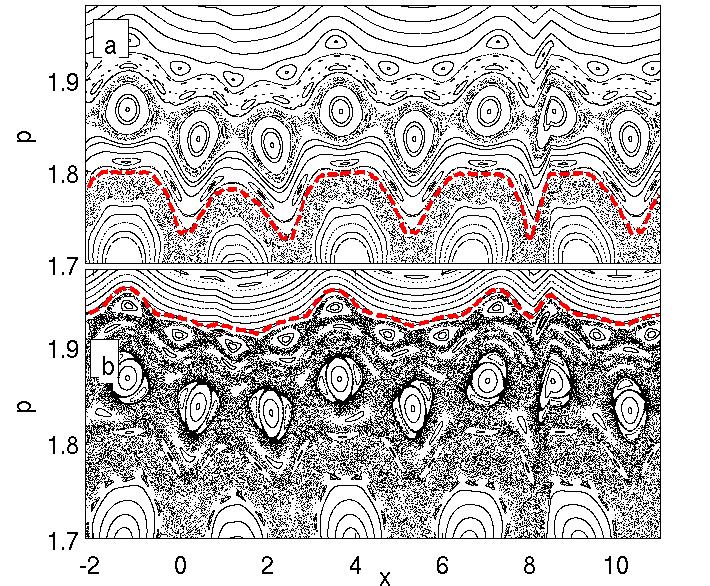}
\caption{\label{fig:fig4} (Color online) Magnification of the Poincar\'{e} surface of section in the upper region of the FISC for two different phase gradients. (a) is the equidistant gradient of period 3 and in (b) the phase of the ``middle'' barrier has been shifted by $\alpha=-0.03\pi/3$. The dashed curves are the FISCs.}
\end{figure}

\paragraph{Perturbed phase gradient.} As the next step, the equidistance of the phase gradient with period 3 is broken up by imposing an additional phase shift $\alpha$ to the ``middle'' barrier. For a relatively large range of $\alpha$ one observes that the PSS remains on a coarse scale unaffected by this perturbation in the sense that the position and the size of the significant resonances stays approximately the same. However, close to the FISC important changes occur. Therefore the upper part of the chaotic sea is magnified in Fig. \ref{fig:fig4} for $\alpha=0$ (a) and $\alpha=-\frac{0.03\pi}{3}$ (b). Obviously, the former stable FISC in the lattice with the equidistant gradient dissolves, when decreasing $\alpha$ from 0 to $-0.03\pi/3$. Thereby, the proportion of the chaotic layer is increased considerably. However, at the position where the FISC is for $\alpha=0$ in Fig. \ref{fig:fig4} (a) a cantorus remains in the PSS for $\alpha=-0.03\pi/3$, which represents a strong barrier, such that the Hamiltonian flow passes through it very slowly. As a consequence it takes long until a trajectory, which has been started with small momentum in the chaotic sea, samples the accessible region of phase space above the cantorus. Consequently, for a finite time this cantorus acts in a similar way like the FISC, i.e. as it would be an impenetrable torus. Yet, in the limit of long simulation time the Hamiltonian flow is able to pass through. Therefore these cantori are usually called ``partial barriers'' and possess a major impact on the transient dynamics and the long-term transport. Thus, one has to be careful, when estimating the transport velocity by simulating an ensemble of particles with small initial momenta in the chaotic sea.

\begin{figure*}
\includegraphics[width=2.0\columnwidth]{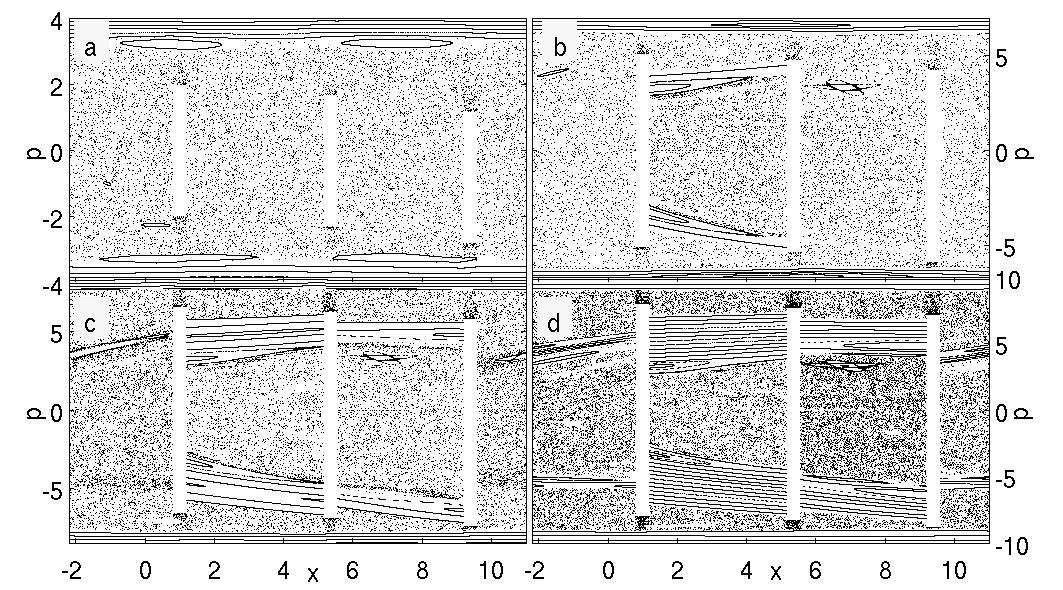}
\caption{\label{fig:fig5} PSS for a lattice with a gradient of period 3 and four different global potential heights. The values are $V_0=2.0$ (a), $V_0=13.0$ (b), $V_0=22.0$ (c) and $V_0=32.0$ (d).}
\end{figure*}

\paragraph{Varying the potential height $V_0$.} Let us now consider the impact of a variation of the global potential height $V_0$ on the phase space properties of a lattice with phase shift period three sequence $\{\ldots,0,\frac{\pi}{10},\frac{3\pi}{10},0,\frac{\pi}{10},\ldots\}$. In Fig. \ref{fig:fig5} the PSS for four different values of $V_0$ is shown. Obviously, the island at $p=0$ being a property of scattering from a single barrier and corresponding to a localized dynamics has disappeared, because the parameters $(V_0,l)$ are chosen such that its central periodic orbit has ceased to exist \cite{Koch:2008}. Still, for not too high values of the potential height $V_0\leq13.0$ the chaotic sea is connected (see Fig. \ref{fig:fig5} (a)). Yet, with increasing $V_0$ another class of localized dynamics arises in the system. As Fig. \ref{fig:fig5} (b) shows, a separated chaotic sea between the first and the second barrier arises. Obviously, this part of phase space is confined by impenetrable tori, i.e. the particles starting in this sea cannot escape from it and are thus trapped between the barriers. (Note that the ``empty barrier'' region in the PSS is not a confinement criterion like the invariant tori.) As the potential height is increased further, this localized chaotic dynamics appears between the other barriers, too. Embedded in these chaotic seas are elliptic islands, e.g. at $x=9.1$, $p=3.2$ in Fig. \ref{fig:fig5} (c). Intuitively, the occurrence of these localized chaotic seas is surprising, because one could think that with increasing potential height the particles just have to acquire more collisions until they are finally fast enough to surpass $V_0$. Even more astonishing is the fact that apparently these regions in phase space do not appear simultaneously between all barriers but successively within increasing $V_0$. Nevertheless, this behavior can be resolved straightforwardly. For small momenta $p^2/2 \ll V_0$ the particles exhibit exclusively reflective collisions, i.e. they cannot penetrate into the barriers. In this regime the dynamics of the Hamiltonian \eqref{eq:ham} is equivalent to the Fermi-Ulam-Model (FUM) of a particle bouncing between two infinitely heavy oscillating walls. In the case of smooth driving laws the phase space of the FUM is not globally stochastic but possesses a FISC at momentum $p_b$, which prevents particles from gaining arbitrarily high momenta \cite{Lich:1992}. Consequently, localized chaos in the case of the driven lattice will occur if the potential height is larger than the kinetic energy associated with this momentum $V_0 \geq p_{b}^2/2$. For momenta less than a certain value $p_s$ the phase space of the FUM is indeed completely chaotic. In the case of the previous setups Figs. \ref{fig:fig3}, \ref{fig:fig4} with $V_0=0.16$ and in Fig. \ref{fig:fig5} (a) the potential height $V_0$ has been chosen smaller than the corresponding kinetic energy $V_0 \leq p_s^2/2$. Consequently, all orbits of the large chaotic sea move diffusively through the lattice in these cases. For intermediate momenta $|p_s|<|p_r|<|p_b|$ the phase space of the FUM possesses a well known resonance structure, whose central periodic orbits correspond to trajectories similar to the one shown in Fig. \ref{fig:traj} (b). When the global potential height is increased, these orbits and their surrounding elliptic islands appear in the PSS between two barriers prior to the blocking tori, since $|p_r|<|p_b|$. Due to the different phase relations the barriers have with respect to each other, these confined chaotic seas do not occur simultaneously between different pairs of barriers but one after another with increasing potential height (see Fig. \ref{fig:fig5} (b)-(d)). Of course, there coexists a chaotic, diffusive dynamics of the particles in the lattice, too.

Let us briefly summarize the different possibilities for trapping particles in one unit cell of the lattice. In the case of small potential heights, which have been used for instance in setup (a) and (b), this can be done by means of the elliptic island at $p=0$ corresponding to trapping in a single barrier. For larger values of $V_0$ this structure is gone, but islands belonging to trapped trajectories between two barriers are present in phase space (see discussion in the previous paragraph). Furthermore, for $V_0\geq13$ there is the possibility of trapping particles in the confined chaotic seas. Finally, we want to remark that for setups with very specifically chosen potential heights and phase gradients it is possible to achieve regular, localized dynamics, which extends over the scattering regions of several barriers. However, this is a case, which requires fine-tuning, since the associated elliptic islands are very tiny and for the parameter regime chosen in our setups this kind of dynamics does not occur.

\section{Transport and localization of particles}\label{ch:trans_loc}

In this section the transport and the localization properties of the chosen setups are studied. As we have seen in section \ref{ch:setup} directed transport can be ruled out for the case of a lattice of uniformly oscillating barriers. Yet, whether the other setups show directed transport cannot be judged prior to a more detailed analysis, because breaking the symmetries (see Eq. \eqref{eq:trans} and Ref. \cite{Flach:2000}) is necessary but not sufficient for the occurrence of directed transport.

\subsection{Transport properties in phase-modulated lattices}

We will deal with the question whether the setups (a)-(c) show a directed flow of particles. In the course of this study the stability of the current against perturbations of the gradient and the diffusion properties in the lattices are discussed, too.

\begin{figure}
\includegraphics[width=\columnwidth]{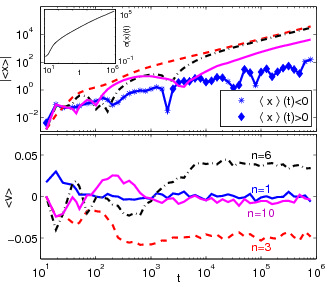}
\caption{\label{fig:trans1} (Color online) In the upper part of the figure the absolute value of the mean position averaged over an ensemble of $10^5$ particles, which has been started with small momentum $p\leq 0.1$ in the chaotic sea as a function of time is shown for various phase periods. For $n=6$ the evolution of the standard deviation has been plotted in the inset. Accordingly, the lower part of the Figure shows the mean velocity as a function of time.}
\end{figure}

\setcounter{paragraph}{0}
\paragraph{Transport properties for constant phase gradients} In the upper part of Fig. \ref{fig:trans1} the evolution in time of the absolute value of the mean position $| \langle x \rangle | (t)$ for several periods of the phase pattern is shown. All particles have been started with small momenta in the chaotic sea close to the origin of the lattice, in order to avoid that some of them are initialized already in a small elliptic island belonging to a ballistic flight. Obviously, for $n=1$ (lattice of uniformly oscillating barriers) no directed transport occurs, as we have expected because in this case the relevant symmetries are not broken (see Sec. \ref{ch:setup}). The ensemble average of the position fluctuates only around $\langle x \rangle =0$. In order to make this visible the data-points with $\langle x \rangle (t)<0$ ($\langle x \rangle (t)>0$) for $n=1$ have been plotted with stars (diamonds). A similar behavior is observed for $\langle v \rangle (t)$ (lower part of Fig. \ref{fig:trans1}). For $n=1$ the mean velocity changes its sign several times correlated with a corresponding behavior of $\langle x \rangle (t)$. Contrary, the other phase periods show clearly directed transport. After an initial transient $t \approx 10^3-10^4$ the mean position grows according to $\langle x \rangle (t) = v_{\text{mean}} \cdot t$, where $v_{\text{mean}}$ is referred to as the \emph{transport velocity} of the system. An asymptotic linear fit to the curves $\langle x \rangle (t)$ yields the mean transport velocities presented in the first row of Table \ref{tab:vel}.
\begin{table}
\begin{tabular}{|c|c|c|c|}\hline
n & 3 & 6 & 10\\
\hline
$v_{\text{mean}}$ & \quad -0.0476 \quad & \quad 0.0384 \quad & \quad -0.0055 \quad \\
\hline
$v_{\mathcal{CS}}$ & \quad -0.0456 \quad & \quad 0.0372 \quad & \quad 0.0201  \quad \\
\hline
\end{tabular}
\caption{\label{tab:vel} Transport velocities for the different phase periods obtained by performing asymptotically a linear fit to the evolution of the ensemble average of the position (first row) and by averaging over the chaotic sea of phase space (second row).}
\end{table}
For $n=3$ and $n=6$ these values for $v_{\text{mean}}$ are consistent with the behavior of the ensemble average of the velocity $\langle v \rangle (t)$ for large times (see lower part of Fig. \ref{fig:trans1}). After the initial transient, $\langle v \rangle (t)$ never changes sign again and fluctuates around $v_{\text{mean}}$. Whether this is also true in the case of the phase period $n=10$ cannot be judged ultimately, since the fluctuations of $\langle v \rangle (t)$ are of the same order of magnitude as $v_{\text{mean}}$. Nevertheless, Fig. \ref{fig:trans1} shows clearly that after the transient $t\approx10^4$ the ensemble average of the position grows linearly with time.

Let us now discuss the question of the origin for the appearance of directed transport. Due to the choice of the initial conditions, the ensemble is localized in the chaotic sea between the FISCs for every point in time. Thus, the convergence of the ensemble average of the velocity $\langle v \rangle (t)$ to a non-zero value can be explained by means of an asymmetry of the chaotic sea with respect to $p=0$, if it is assumed additionally that the dynamics is ergodic with a uniform invariant density \cite{Schanz:2005}. Again, it is important to emphasize that this desymmetrization has to occur in phase space and not only for a certain PSS, which depends on the phase of driving at the moment of the snapshot. For Hamiltonian system with mixed phase space the proof that the dynamics is ergodic in the chaotic subsets is still an open problem \cite{Buni:2008} and has been achieved only for very special systems \cite{Buni:2001}. Nevertheless, if we assume ergodicity of the dynamics in the chaotic sea, then the phase space average of the velocity $v_{\mathcal{CS}}$ over the chaotic sea should coincide with $v_{\text{mean}}$. $v_{\mathcal{CS}}$ can be calculated by
\begin{equation}\label{eq:vmean1}
v_{\mathcal{CS}}=\frac{1}{\Omega_{\mathcal{CS}}}\int \limits_\mathcal{CS} \frac{p}{m} \, d\Gamma, \quad \text{with} \enspace d\Gamma=dx\,dp\,d\xi,
\end{equation}
where $\Omega_{\mathcal{CS}}$ is the phase space volume of the chaotic sea and $\xi$ is the phase of the driving. Due to the time-periodicity of the Hamiltonian $H(x,p,t+2\pi)=H(x,p,t)$, the volume $\Omega_{\mathcal{CS}}$ can be written as $\Omega_{\mathcal{CS}}=2\pi \cdot A_{\mathcal{CS}}$ with $A_{\mathcal{CS}}$ being the area of the chaotic sea in the PSS. Performing the integration over $x$ and $p$ yields hence the mean velocity of the chaotic sea in the PSS, which is determined for a fixed $\xi$ by dividing the PSS into small rectangles and averaging over the cells that get visited by a single long chaotic trajectory. This yields a function $v_{\mathcal{PSS}}(\xi)$, whose convergence is checked by enlarging the grid of the PSS. Finally, the mean velocity of the chaotic sea of the complete phase space is obtained by averaging over the phases of the PSS, i.e.
\begin{equation}\label{eq:vmean2}
v_{\mathcal{CS}}=\frac{1}{2\pi}\int \limits_0^{2\pi} v_{\mathcal{PSS}}(\xi) d\xi.
\end{equation}
Of course this scheme is equivalent to the procedure presented in \cite{Schanz:2005}. We remark, that for our system it turns out to be more efficient to evaluate numerically the integral \eqref{eq:vmean1} than applying the sum formula of \cite{Schanz:2005}, where all the winding numbers and areas of the significant elliptic resonances in the PSS have to be determined. Furthermore, even for very long simulation times ($t \geq 10^9$) a ``drift'' of a chaotic trajectory beyond an intact KAM-torus by accumulating numerical errors has never been observed in our system. For the lattice of uniformly oscillating barriers, our procedure gives $v_{\mathcal{CS}}=0$, as expected. In the case of other phase periods, i.e. different from one Eq. \eqref{eq:vmean2} yields non-zero values for $v_{\mathcal{CS}}$, which are summarized in the second row of Table \ref{tab:vel}. For $n=3$ and $n=6$ the velocities obtained by performing asymptotically a linear fit to the mean position and averaging over the chaotic sea coincide very well. However, for $n=10$ Eq. \eqref{eq:vmean1} predicts that the transport should be actually in the opposite direction than it is observed (see Fig. \ref{fig:trans1}). This contradiction can be resolved straightforwardly. According to our discussion of the PSS, there are cantori in the chaotic parts of phase space. Depending on the system's parameters like phase period or potential height, the flux across them can become arbitrarily small and therefore these cantori restrict trajectories to certain subparts of phase space even for long simulation times. In the previous section their impact on the appearance of ballistic flights has been discussed. Yet, as we have indicated already in the discussion of the setup with the perturbed phase period three, there is of course the opposite effect, too. For trajectories starting with small momenta like our initial ensembles, the cantori prevent the particles even for long times from sampling the (transporting) phase space beyond the cantori. Indeed the surfaces of section of phase period $n=10$ for $t<10^6$ show that the region of the chaotic sea especially close to the FISC for positive momenta has not been visited by a single trajectory at all. Since this part of phase space corresponds to a positively valued drift, a transient transport in the opposite direction occurs in our case. In order to verify this explanation, a smaller initial ensemble has been simulated for a longer simulation time. For this simulation a zero crossing of $\langle x \rangle (t)$ is observed at $t_{\text{cr}} \approx 1.3 \cdot 10^7$. After the zero crossing the mean position $\langle x \rangle (t)$ grows in the positive $x$-direction, which is predicted by equation \eqref{eq:vmean1}. Moreover, for $t>t_{\text{cr}}$ the region in the PSS close to the FISC for positive momenta gets sequentially visited by trajectories.

According to the above, the origin for the occurrence of directed transport is simply the desymmetrization of the chaotic sea with respect to $p=0$, i.e. for the phase periods larger than three the phase space volume of the chaotic sea with $p>0$ and accordingly $p<0$ are not equal anymore. An important manifestation of this asymmetry is that the areas of elliptic islands with $w=r/s$ and $w=-r/s$ are different in the PSS. Usually, this phase space asymmetry is achieved by applying a biharmonic driving law to the static system \cite{Gommers:2005,Salger:2009,Flach:2000,Denisov:2001,Denisov:2002,Schanz:2005}, such that the potential $V(x,t)$ of the Hamiltonian breaks both symmetries derived in \cite{Flach:2000}. However, from the discussion of the FISC in the previous section we know that a kinematic approach to the single barrier dynamics explains already the asymmetry of the chaotic sea of the phase space with respect to $p=0$, because the FISC, depending on the direction of the propagation, is located in different ranges of the velocity. For the lattice with phase-modulated harmonically driven barriers this simple consideration, independently of the phase period, is not enough to explain the asymmetry.

\paragraph{Diffusion properties for constant phase gradients} In Sec. \ref{ch:pss} it has been shown that every barrier possesses an elliptic island at $p=0$ for the parameter values $(V_0,l)$ chosen for setup (a). Accordingly, the particles in the chaotic sea, which move diffusively through the lattice, can become sticky to this structure. They therefore remain at the same spatial location for many periods of the driving. On the other hand the particles can perform ballistic flights. These events originate either from phases of motion during which the particles are confined by partial barriers to regions of phase space with non-zero average velocity or they are the result of stickiness to elliptic islands corresponding to periodic orbits, which travel through the lattice in the direction of their initial momentum. Usually, strong partial barriers are close to the FISCs, where transporting islands are found, too. Consequently, both effects reinforce each other in this part of phase space, i.e. trajectories which are confined by a partial barrier and get additionally sticky to elliptic islands contribute to the longest ballistic flights in the system.

\begin{figure}
\includegraphics[width=\columnwidth]{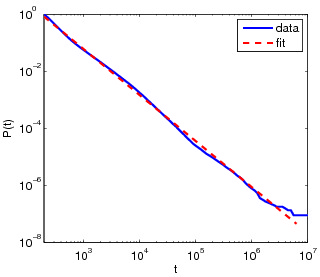}
\caption{\label{fig:levi} Cumulative distribution function $P(t)$ for the length of ballistic flights (phase period $n=6$). Fitting a power law to the distribution yields an exponent of $\mu=1.6$.}
\end{figure}

These trajectories can be detected numerically very easily. We look simply for time intervals with a minimum length, where the sign of the velocity does not change upon interaction between the particle and the barriers, i.e. we search for ballistic flights. The corresponding probability distribution of the lengths of ballistic flights obeys a power law $p(t) \sim t^{-\nu}$. However, in order to reduce statistical noise \cite{Newmann:2005}, we do not plot a simple histogram of the data, but calculate the cumulative probability distribution function $P(t)$, which is defined by
\begin{equation}\label{eq:pl}
P(t)=\int_t^{\infty} p(t') \, dt'.
\end{equation}
Fig. \ref{fig:levi} shows $P(t)$ in a double-logarithmic plot. Clearly, there is a power-law behavior of $P(t) \sim t^{-\mu}$ over several decades. The exponent can be estimated by a least-square fit using a power law, which gives $\mu=1.6$. Thus, due to equation \eqref{eq:pl} the probability distribution $p(t)$ follows a power law with the exponent $\nu=\mu+1=2.6$. From continuous-time-random-walk (CTRW) theory \cite{Montroll:1965,Denisov:2001,Denisov:2002} it is known that this interplay between ballistic flights and waiting times should yield anomalous diffusion in configuration space. In the inset of Fig. \ref{fig:trans1} the variance $\sigma(x)(t)$ for $n=6$ is shown, which obviously grows for large times according to $\sigma(x)(t) \sim t^{\gamma}$. Performing an asymptotic power-law fit yields for the exponent $\gamma \approx 0.65$. CTRW theory predicts that between $\gamma$ and $\mu$ the relation $2\gamma=3-\mu$ \cite{Zumofen:1993,Klafter:1994} holds. For our numerical data this is quite well fulfilled. A similar behavior is found for the other phase periods, too. In the asymptotic time limit the variance follows a power-law and the exponent is always between $0.5$ and $1$, i.e. the system shows universally superdiffusion in configuration space.

\begin{figure}
\includegraphics[width=\columnwidth]{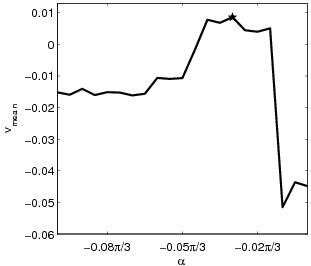}
\caption{\label{fig:trans_pert} Transport velocity $v_{\text{mean}}$, as a function of the phase shift $\alpha=-\frac{0.1\pi}{3}\ldots 0$ of the ``middle'' barrier. With a star we have marked the setup, which we have chosen in section \ref{ch:pss} to produce the PSS shown in Fig. \ref{fig:fig4} (b).}
\end{figure}

\paragraph{Transport properties for the perburbed / broken phase gradient}Now we turn to the discussion of the transport properties of the lattice, where the phase period three with equidistant gradient is perturbed. Fig. \ref{fig:trans_pert} shows the transport velocity of the system as a function of the phase shift $\alpha$ of the ``middle'' barrier. For these setups and all the upcoming ones the transport velocity has been calculated exclusively by evaluating numerically Eq. \eqref{eq:vmean1}, because this is much more efficient in terms of computational time than simulating the long term dynamics of a whole ensemble. The star in Fig. \ref{fig:trans_pert} marks the setup with $\alpha=-\frac{0.03\pi}{3}$ for which one obtains the PSS shown in Fig. \ref{fig:fig4} (b). Using this example one can again demonstrate very vivid the impact of the tuning of the asymmetric phase space on the direction of the transport. The PSS of the setup with equidistant phase period three, i.e. $\alpha=0$, shown in Fig. \ref{fig:fig4} (a) corresponds to the system with transport velocity $v_{\mathcal{CS}}(\alpha=0)=-0.0456$. By perturbing this equidistance of the gradient the formerly stable FISC dissolves, which makes parts of the phase space volume with larger positive momenta accessible to chaotic trajectories. At the same time, the remaining phase space, in particular the part with $p<0$, remains basically unaffected, i.e. overall the phase space volume of the chaotic sea with $p>0$ increases significantly. Therefore, the phase space average of the velocity over the chaotic sea in equation \eqref{eq:vmean1} and thus the direction of the transport changes its sign. For some values of $\alpha$ the directed transport vanishes, i.e. $v_{\mathcal{CS}}(\alpha)=0$. Yet, this does not imply that a symmetry of type \eqref{eq:trans} is present. In fact both symmetries are broken. Nevertheless, the phase space volumes with $p>0$ and $p<0$ of the chaotic sea are accidentally equal for some values of $\alpha$.

Finally, we consider the transport properties of the setup with variable potential height and fixed phase gradient of period three $\{\ldots,0,\frac{\pi}{10},\frac{3\pi}{10},0,\frac{\pi}{10},\frac{3\pi}{10},\ldots \}$. By evaluating Eq. \eqref{eq:vmean1} the transport velocities depending on the global potential heights are calculated and summarized in Table \ref{tab:velV0}. Similarly to the previous setup, the directed transport can be tuned by varying a parameter, which is now the potential height.

\subsection{Localization and trapping}

\begin{figure}
\includegraphics[width=\columnwidth]{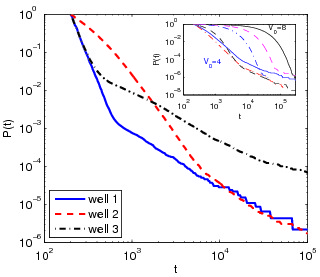}
\caption{\label{fig:dwell} (Color online) Cumulative distribution function $P(t)$ of the dwell time for the three different wells with fixed potential height $V_0=4$ for setup (c). In the inset $P(t)$ belonging to the second well for different values of $V_0=4\ldots8$ is shown.}
\end{figure}

In this section the three different possibilities for trapping, which have been summarized in Sec. \ref{ch:pss1}, are discussed in more detail. Especially the impact of the potential height is addressed, because the localization of particles is most sensitive to this parameter.
\setcounter{paragraph}{0}
\paragraph{Localization properties for a broken phase gradient} In the course of the discussion of the diffusion properties it has been shown that the trajectories, which contribute to the directed current in the lattice, can still obey for long time phases of motion during which they are localized in the scattering region of one barrier. For small potential heights these events originate from stickiness to the elliptic island at $p=0$, whose properties are defined exclusively by $(V_0,l)$. Since these parameters are equal for all barriers, i.e. a single barrier property \cite{Koch:2008}, the stickiness of trajectories to this structure is identical independently of the phase periods: The particles get localized in the scattering region of the first, second etc. barrier on average for the same time. In the case of very large potential heights, e.g. $V_0=32$ for setup (c), there are confined chaotic seas between the barriers. Yet, the particles belonging to the directed flow cannot enter these regions in phase space, because these parts are separated by impenetrable tori. In the following we will show that it is possible for certain parameter values to obtain a dynamics that exhibits phases of trapped motion in certain wells, i.e. in between certain definite barriers, of the driven unit cell of the lattice. To this end we look in the case of setup (c) for time intervals during which the particles, that move diffusively through the lattice, are localized between two barriers. Fig. \ref{fig:dwell} shows for $V_0=4$ the cumulative probability distribution of dwell times $P(t)$ for particles between two consecutive barriers in one spatial unit cell, i.e.  $i=-1$ to $i=0$, $i=0$ to $i=1$ and $i=1$ to $i=2$, which will be labeled as well one, two and three, respectively. For this value of the potential height (see Sec. \ref{ch:pss1}) the elliptic island at $p=0$ has disappeared and thus there is no stickiness to this regular structure anymore. Furthermore, no trapped chaotic dynamics between barriers is possible. Instead all chaotic trajectories in the large chaotic sea move diffusively through the lattice. Obviously, $P(t)$ is not equal for the different wells. A particle, which has been localized in the first well is trapped with less probability (for a time span $t\geq10^3$) compared to a particle trapped in the third well. By recalling the discussion of the PSS for this values of $V_0$ in Sec. \ref{ch:pss1} this behavior can be resolved. Accordingly, with increasing potential height the particles between two barriers are able to probe parts of the phase space of the corresponding FUM, which are located at higher momentum. However, the phase space of the FUM in general is very sensitive to the phase relations the walls have with respect to each other. Not only the position in momentum space of the FISC $p_b$, but also both the momentum $p_s$ below which the FUM is completely chaotic and the momentum of elliptic islands belonging to trajectories shown in Fig. \ref{fig:traj} (b) depend critically on the phase relations between the barriers. Due to the non-equidistant gradient these phase relations are not equal, i.e. the barriers belonging to different wells possess different relative phases with respect to each other. Thus, for a fixed potential height it is possible that in a certain well only the chaotic part of the corresponding FUM's phase space is accessible to the particles, whereas in another well already elliptic islands embedded in the chaotic sea are present.
\begin{table}
\begin{tabular}{|c|c|c|c|c|}\hline
$V_0$ & 2 & 13 & 22 & 32\\
\hline
\quad $v_{\mathcal{CS}}$ \quad & \quad 0.0221 \quad & \quad 0.0404 \quad & \quad 0.0088 \quad & \quad 0.1201 \quad\\
\hline
\end{tabular}
\caption{\label{tab:velV0} Transport velocities for different global potential heights $V_0$ obtained by averaging over the chaotic sea of phase space.}
\end{table}
Escape rates from these different parts of phase space are understandably not equal, because stickiness to regular structure leads to phases of motion during which particles are trapped for long times between two barriers. By changing the phase gradient for a fixed potential height, it is possible to ``engineer the phase space of the lattice in such a way'' that the events with very long dwell times take place only in specific wells. Of course, trajectories can also stay for a comparatively long time in a well, which does not contain any elliptic island at all, which is due to the so-called \emph{low velocity peaks} \cite{Papa:2001}. Particles, that are slightly faster than the barrier at a collision, get decelerated to a very small velocity $v_{\epsilon}$ but still leave the scattering region without experiencing a second collision. Afterwards they travel the spatial distance between the scattering regions of two barriers, which yields for the dwell time $t_d=\frac{D-l-2C}{v_{\epsilon}}$. Yet, the portion of trajectories, which are ``trapped'' according to this mechanism, is negligible. For $V_0=4$ all wells contain elliptic islands. In fact, the characteristic asymptotic power-law behavior $P(t) \sim t^{-\mu}$ of the cumulative probability distributions of dwell times is a hallmark of the associated stickiness to this regular structure in phase space. Generally, whenever a region in phase space contains elliptic islands, the occurrence of long algebraic tails in the escape rate is generic \cite{Meiss:1992}. Indeed, we observe that the exponent of the probability distribution $p(t)\sim t^{-\nu}$, which is related to $P(t)$ through equation \eqref{eq:pl}, is always larger than two. Similarly to the previous setups, the dynamics shows the characteristic interplay between ballistic flights and waiting times. Accordingly, we find that there is superdiffusion in coordinate space even for very large potential heights.

\paragraph{Impact of potential height on localization properties}Finally, let us discuss the variation of $V_0$ influences $P(t)$ for a given well. In the inset of Fig. \ref{fig:dwell} the cumulative probability distribution of the dwell time in the second well for various values of $V_0$ between 4 and 8 is shown. With increasing $V_0$ this is the first well with a localized chaotic sea (see Fig. \ref{fig:fig5} (b)). Especially for small and intermediate dwell times ($10^2<t<3 \cdot 10^3$) the function $P(t)$ changes significantly. $P(t)$ decreases less rapidly for larger values of $V_0$, which is understandable, because with increasing potential height the particles need to accumulate more collisions in order to become fast enough to surpass the potential. Still, by doubling $V_0$ the probability for the particles to be trapped in the well for dwell times longer than $10^4$ grows up five orders of magnitude. Furthermore, the PSSs of trajectories with dwell times in this range show that beside the particles, which are sticky to elliptic islands, a great portion still obeys chaotic dynamics. In order to answer the question how chaotic particles, which move diffusively through the lattice, can be trapped between two barriers for such a long time, we have to exploit one more time the FUM. Close to the FISC of the FUM are again cantori with small flux across them. Depending on the properties of these partial barriers the particles obeying chaotic dynamics are prevented for a significant time from sampling the phase space of the FUM, which is located at high momenta close to the FISC and associated to the kinetic energy a trajectory has to gain in order to leave the well. Consequently, once the potential height is chosen such that this cantori are stabilized, the escape rate from the region between barriers is lowered significantly.

\section{Conclusions and outlook}\label{ch:sum}

We have explored the dynamics of non-interacting particles in one-dimensional, phase-modulated driven lattices. Depending on the parameters like the potential height and phase period the phase space of the system has been analyzed in detail. The impact of cantori on the transient transport properties, as well as the occurrence of ballistic flights due to the dynamical confinement of trajectories to volumes in phase space with non-zero velocity, has been studied. Depending on the phase period a directed current is observed, whose occurrence has been traced back to the desymmetrization of the chaotic sea \cite{Schanz:2001,Schanz:2005,Dittrich:2000}. Commonly, this asymmetry of the phase space is achieved by breaking two established spatiotemporal symmetries \cite{Flach:2000} with an appropriately chosen biharmonic driving \cite{Gommers:2005,Salger:2009,Flach:2000,Denisov:2001,Denisov:2002,Schanz:2005}. For this driving law a kinematic argument considering the single barrier dynamics is sufficient to explain the occurrence of a desymmetrization of the extended system's phase space, i.e. each barrier is transporting itself and the lattice inherits this property. In this work it has been shown that a simple harmonic driving law together with a local symmetry breaking implemented by local phase shifts is sufficient to generate a flow of particles although each individual barrier is non-transporting. Consequently, the occurrence of the particle current is in this case a collective phenomena of the complete lattice. The direction and the magnitude of the transport can be tuned by the parameters of the system like the phase period or the potential height.

Of course the most simplest way to revert the direction of the particle current is by inverting the gradient. With increasing potential height elliptic islands belonging to a trapping dynamics between the barriers arise. Their properties like size and position are influenced by the phase relation of the barriers with respect to each other. Accordingly, the phase space can be locally manipulated in a controllable manner, such that each well possesses its own characteristic escape rate. Furthermore, it has been shown that the system shows universally superdiffusion in coordinate space. In retrospect this occurrence of anomalous diffusion is understandable, because its prerequisites are always present in phase space. Independent of the system's parameters like potential height, phase period etc. the phase space is mixed. Thus, it possesses the hierarchy of elliptic islands surrounded by subislands. Events of stickiness to this regular structure yield either ballistic flights or waiting times. In both cases we have found that the corresponding probability distributions follow asymptotically typical power-laws, which give rise to anomalous diffusion and can be modeled by continuous time random walk theory. A major advantage of the presented setup compared to other driven lattice systems is that the driving can be locally adjusted to engineer the phase space. Thereby, the desired transport and localization properties can be achieved much more simply than by applying just another driving law. Furthermore, the local driving laws offers the possibility of building up ``blocks of sublattices'', such that a current of particles can be manipulated.

\section*{Acknowledgments}
Part of this work was financially supported in the framework of the excellence initiative of the federal and state government of Germany, i.e. the FRONTIER program of the university of Heidelberg. FKD thanks the Heidelberg Graduate School of Fundamental Physics for financial support in the framework of a visit to the university oh Heidelberg. The authors acknowledge helpful discussions with Florian Koch concerning the technical and numerical aspects of the present work.

\appendix

\section{Determining the position of periodic orbits in the PSS with few collisions}\label{ap:po_few}

In the following we describe briefly our method to determine the position of the periodic orbits with winding number $w=r$ in the PSS using the example of $w=2$. For this trajectory the initial velocity $v_0$ and the particle's velocity between the first and the second barrier $v_2$ (see Fig. \ref{fig:traj} (a)) are defined by the condition that the phases at collision with the first and third barrier are the same. In the case of the uniformly oscillating lattice $\{\varphi_i=\text{const.} \, \forall i \}$, this yields after some algebra
\begin{gather}
v_0=\frac{L-l+2 \cos(\xi_1)}{\pi+2\xi_1},\\
v_2=\frac{L-l-2 \cos(\xi_2)}{\pi-2\xi_2}.
\end{gather}
Due to the symmetry of the phases $\xi$ upon collision, the particle's velocity after the first and the third collision with one of the barriers' edges are equal, i.e. $v_1=v_3$. Furthermore, the spatial distance between the collisions at phase $\xi_1$ and $\xi_2$ has to be covered by the particle during $\Delta t=\xi_2-\xi_1$. Both conditions yield a system of two nonlinear equations
\begin{widetext}
\begin{equation}\label{eq:imp}
\begin{split}
f_1(\xi_1,\xi_2)&=\sin(\xi_1)-\sin(\xi_2)-\sqrt{\left(v_0(\xi_1)+\sin(\xi_1)\right)^2-2 V_0}+\sqrt{\left(v_2(\xi_2)+\sin(\xi_2)\right)^2-2 V_0}=0,\\
f_2(\xi_1,\xi_2)&=\cos(\xi_1)-\cos(\xi_2)+\left(\xi_2-\xi_1\right)\left(\sqrt{\left(v_0(\xi_1)+\sin(\xi_1)\right)^2-2 V_0}-\sin(\xi_1)\right)=0.
\end{split}
\end{equation}
\end{widetext}
Solving equation \eqref{eq:imp} for $\xi_1$ and $\xi_2$, the position of the elliptic island in the PSS can be calculated straightforwardly. In general, one can proceed similarly for setups with larger phase periods and for every resonance. A system of equations will therefore result
\begin{equation}\label{eq:imp_large}
\{f_i(\xi_1,\xi_2,\cdots,\xi_k)=0, \enspace i=1\ldots k\}.
\end{equation}
We remark that it is difficult to generalize this scheme to periodic orbits with many collisions, since the approach to set up the system of equations and the roots corresponding to physical solutions depend on the specific symmetry properties of the phases upon collisions of the central periodic orbit belonging to the resonance.

\section{Finding periodic orbits with arbitrary winding number}\label{ap:po_many}

For periodic orbits with many collisions, which are needed for the calculation of the flux through a cantorus, the scheme presented in appendix \ref{ap:po_few} is obviously not feasible anymore. In this case it makes more sense to make use of a variational method, which will be presented in the following. Since the particles move ballistically, the trajectories in the lattice belonging to the periodic orbits in the PSS are completely defined by the moments of the collisions $t_k$ with the barriers' edges. Thus the action \eqref{eq:action} of such a periodic orbit can be rewritten as
\begin{equation}\label{eq:action_re}
W_{r/s}=\sum_{k=1}^N W_k(t_k,t_{k+1}),
\end{equation}
i.e. the action between successive collisions is added piecewise. $N$ is the total number of collisions. Now we restrict ourselves to periodic orbits with winding number $w=r/s$. Firstly, this yields that in equation \eqref{eq:action_re} $t_{N+1}=t_1+2\pi s$. Secondly, $r$ specifies how many spatial unit cells a trajectory travels in the lattice, at which the velocity never changes sign. Accordingly, the number of collisions $N$, the phase period $n$ and $r$ are related through $N=2\cdot n \cdot r$. Inserting the Lagrangian of the system \eqref{eq:lagrangian} in equation \eqref{eq:action_re} gives
\begin{widetext}
\begin{equation}\label{eq:action_re1}
W_k(t_k,t_{k+1})=
\begin{cases}\frac{m \left( x(t_{k+1})-x(t_k) \right)^2}{2 \left(t_{k+1}-t_k\right)} & \text{particle between barriers}\\
\frac{m \left( x(t_{k+1})-x(t_k) \right)^2}{2 \left(t_{k+1}-t_k\right)}-V_0 \left(t_{k+1}-t_k\right) & \text{particle in barrier}\end{cases}
\end{equation}
\end{widetext}
Of course $x(t_k)$ equals with the position of one of the barriers' edges at a collision, which depends on the driving law $f_i(t)$ at the {\it i}-th site. For particles in a barrier the spatial distance between the successive collisions is $x(t_{k+1}-x(t_k))=l+f_i(t_{k+1})-f_i(t_k)$ and accordingly $x(t_{k+1}-x(t_k))=D-l+f_{i+1}(t_{k+1})-f_i(t_k)$ between the barriers. Setting $m=1$ the action becomes
\begin{widetext}
\begin{equation}
W_k(t_k,t_{k+1})=
\begin{cases}\frac{\left( D-l+f_{i+1}(t_{k+1})-f_i(t_k) \right)^2}{2 \left(t_{k+1}-t_k\right)} & \text{particle between barriers}\\
\frac{\left( l+f_i(t_{k+1})-f_i(t_k) \right)^2}{2 \left(t_{k+1}-t_k\right)}-V_0 \left(t_{k+1}-t_k\right) & \text{particle in barrier}\end{cases}
\end{equation}
\end{widetext}
We search for trajectories for which the action is extremal. The action gradient vector is $\nabla W=\left( \partial{W}/\partial{t_1},\ldots,\partial{W}/\partial{t_N}\right)^T$ with
\begin{equation}
\frac{\partial{W}}{\partial{t_i}}=\frac{\partial{W_{i-1}(t_{i-1},t_i)}}{\partial{t_i}}+
\frac{\partial{W_{i}(t_{i},t_{i+1})}}{\partial{t_i}}
\end{equation}
For finding roots of $\nabla W$ a multidimensional Newton scheme has been applied. Therefore, the derivative matrix (``Hessian'') of the action gradient is needed. It is a cyclic, tridiagonal matrix of the second derivatives of $W_i$, whose rank increases with the period {\it s} of the orbit.  To get an initial starting point we start from the integrable limit ($V_0=0$) and trace the periodic orbits as the perturbation, i.e. $V_0$, is increased. Finally, the stability of the obtained periodic orbit can be determined via the eigenvalues of the Hessian. In the case of the minimizing orbit it has only positive eigenvalues, whereas for the minimax orbit there is a single negative eigenvalue. For orbits with not too high periods ($s<300$) the results of this scheme have been compared additionally to other globally convergent methods \cite{Schmelcher:1997,Schmelcher:1998} and we have found good agreement.


\begin{thebibliography}{9}
\bibitem{Gavrila:1992} M. Gavrila, {\it Atoms in Intense Laser Fields}, (Academic, San Diego, 1992).
\bibitem{Delone:1995} N. B. Delone and V. P. Krainov, {\it Multiphoton Processes in Atoms}, (Springer, Heidelberg, 1995).
\bibitem{Krausz:2000} T. Brabec and F. Krausz, Rev. Mod. Phys. {\bf 72}, 545 (2000).
\bibitem{Reimann:2002} P. Reimann, Phys. Rep. {\bf 361}, 57 (2002).
\bibitem{Astumian:2002} R. D. Astumian and P. H\"{a}nggi, Phys. Today {\bf 55}, 33 (2002).
\bibitem{Goychuck:2001} I. Goychuk and P. H\"{a}nggi, J. Phys. Chem. B {\bf 105}, 6642 (2001).
\bibitem{Haenggi:1996} P. Jung, J. G. Kissner, and P. H\"{a}nggi, Phys. Rev. Lett. {\bf 76}, 3436 (1996).
\bibitem{Mateos:2000} J. L. Mateos, Phys. Rev. Lett. {\bf 84}, 258 (2000).
\bibitem{Mateos:2003} J. L. Mateos, Physica A {\bf 325}, 92 (2003).
\bibitem{Linke:1999} H. Linke {\it et al.}, Science {\bf 286}, 2314 (1999).
\bibitem{Majer:2003} J. B. Majer, J. Peguiron, M. Grifoni, M. Tusveld, and J. E. Mooij, Phys. Rev. Lett. {\bf 90}, 056802 (2003).
\bibitem{Schiavoni:2003} M. Schiavoni, L. Sanchez-Palencia, F. Renzoni, and G. Grynberg, Phys. Rev. Lett. {\bf 90}, 094101 (2003).
\bibitem{Gommers:2005} R. Gommers, S. Bergamini, and F. Renzoni, Phys. Rev. Lett. {\bf 95}, 073003 (2005).
\bibitem{Salger:2009} T. Salger {\it et al.}, Science {\bf 27}, 1241 (2009).
\bibitem{Flach:2000} S. Flach, O. Yevtushenko, and Y. Zolotaryuk, Phys. Rev. Lett. {\bf 84}, 2358 (2000).
\bibitem{Monteiro:2002} T. S. Monteiro, P. A. Dando, N. A. C. Hutchings, and M. R. Isherwood, Phys. Rev. Lett. {\bf 89}, 194102 (2002).
\bibitem{Hutchings:2004} N. A. C. Hutchings, M. R. Isherwood, T. Jonckheere, and T. S. Monteiro, Phys. Rev. E {\bf 70}, 036205 (2004).
\bibitem{Brumer:2006} J. Gong and P. Brumer, Phys. Rev. Lett. {\bf 97}, 240602 (2006).
\bibitem{Gong:2004} J. Gong and P. Brumer, Phys. Rev. E {\bf 70}, 016202 (2004).
\bibitem{Casati:2007} L. Cavallasca, R. Artuso and G. Casati, Phys. Rev. E {\bf 75}, 066213 (2007).
\bibitem{Denisov:2001} S. Denisov and S. Flach, Phys. Rev. E {\bf 64}, 056236 (2001).
\bibitem{Denisov:2002} S. Denisov, J. Klafter, M. Urbakh, and S. Flach, Physica D {\bf 170}, 131 (2002).
\bibitem{Schanz:2005} H. Schanz, T. Dittrich, and R. Ketzmerick, Phys. Rev. E {\bf 71}, 026228 (2005).
\bibitem{Schanz:2001} H. Schanz, M. F. Otto, R. Ketzmerick, and T. Dittrich, Phys. Rev. Lett. {\bf 87}, 070601 (2001).
\bibitem{Dittrich:2000} T. Dittrich, R. Ketzmerick, M. F. Otto, and H. Schanz, Ann. Phys. {\bf 9}, 755 (2000).
\bibitem{Denisov:2007} S. Denisov, L. Morales-Molina, S. Flach, and P. H\"{a}nggi, Phys. Rev. A {\bf 75}, 063424 (2007).
\bibitem{Denisov:2006} S. Denisov, S. Flach, and P. H\"{a}nggi, Europhys. Lett. {\bf 74}, 588 (2006).
\bibitem{Koch:2008} F. R. N. Koch, F. Lenz, C. Petri, F.K. Diakonos, and P. Schmelcher, Phys. Rev. E {\bf 78}, 056204 (2008).
\bibitem{Meiss:1992} J. D. Meiss, Rev. Mod. Phys. {\bf 64}, 795 (1992).
\bibitem{Mather:1986} J. N. Mather, IHES Publ. Math., {\bf 63}, 153 (1986).
\bibitem{MacKay:1983} R. S. MacKay, J. D. Meiss, and I. C. Percival, Physica D {\bf 13}, 55 (1984).
\bibitem{Lich:1992} A.J. Lichtenberg and M.A. Liebermann, {\it Regular and Chaotic Dynamics}, Appl. Math. Sci. {\bf 38}, (Springer Verlag, New York, 1992) Vol. 38.
\bibitem{Buni:2008} L. A. Bunimovich, Nonlinearity {\bf 21}, T13-T17 (2008).
\bibitem{Buni:2001} L. A. Bunimovich, Chaos {\bf 11}, 802 (2001).
\bibitem{Newmann:2005} M. E. J. Newmann, Contemporary Physics {\bf 46}, 323 (2005).
\bibitem{Montroll:1965} E. Montroll and G. H. Weiss, J. Math. Phys. {\bf 6}, 167 (1965).
\bibitem{Zumofen:1993} G. Zumofen, J. Klafter, and A. Blumen, Phys. Rev. E {\bf 47}, 2183 (1993).
\bibitem{Klafter:1994} J. Klafter and G. Zumofen, Phys. Rev. E {\bf 49}, 4873 (1994).
\bibitem{Papa:2001} P. K. Papachristou, F. K. Diakonos, E. Mavrommatis, and V.
Constantoudis, Phys. Rev. E {\bf 64}, 016205 (2001).
\bibitem{Schmelcher:1997} P. Schmelcher and F. K. Diakonos, Phys. Rev. Lett. {\bf 78}, 4733 (1997).
\bibitem{Schmelcher:1998} P. Schmelcher and F. K. Diakonos, Phys. Rev. E {\bf 57}, 2739 (1998).
\end{thebibliography}
\end{document}